\documentclass[12pt]{article}
\usepackage{amsmath}
\usepackage{amssymb}
\usepackage{graphicx}
\usepackage{geometry}
\usepackage{setspace}
\usepackage{tabularx} % Required package for the X column type
\usepackage{ragged2e}
\usepackage{booktabs}
\usepackage{multirow}
\usepackage{caption}
\title{ Decisions of Public Goods Game Through the lens of Game Theory}
\author{Yash Prajapati \\ Department of Mathematics \& Economics \\ University of Kansas}
\date{November 9, 2025}

\begin{document}

\maketitle

\begin{abstract}
    This paper examines public goods and evaluates the mechanism through the game theory. Public goods are characterized by nonexclusivity and nonrivalry and this creates fundamental challenges for allocation. We analyze why competitive markets undersupply public goods by deriving the inefficiency formally through Nash equilibrium. The paper evaluates theoretical solutions including Lindahl pricing, Clarke-Groves mechanisms, and voting schemes. The paper will cover their efficiency properties and practical limitations. We show how strategic interaction leads to free-riding behavior using roommates dilemma and other examples. We also cover why a large household lives in messy conditions not because individuals are lazy,
but because they are rational players in a Nash equilibrium. We also examine voting mechanisms, the median voter theorem, and recent developments in truth-revealing mechanisms
\end{abstract}

\section{Introduction}

A \textbf{public good} is defined by two critical properties: \textit{nonrivalry} and \textit{nonexclusivity}. Nonrivalry means that the marginal cost of serving an additional user is zero, enabling multiple individuals to consume the same unit simultaneously without diminishing others' consumption. Consumption of one more burger equires that various resources be devoted to its production. Nonexclusivity implies that no one can be prevented from consuming the good, even if they do not pay for it. Classic examples include national defense, radio broadcasts, street lighting, and environmental protection.

The concepts of nonexclusion and nonrivalry are in some ways related. Many nonexclusive
goods are also nonrival. National defense and mosquito control are two examples of goods
for which exclusion is not possible and additional consumption takes place at zero marginal cost. Many other instances might be suggested. The concepts, are not identical: Some goods may possess one property but not the other. For example, it is impossible (or at least very costly) to exclude some fishing boats from ocean fisheries, yet the arrival of another boat clearly imposes social costs in the form of a reduced catch for all con cerned. Similarly, use of a bridge during off-peak hours may be nonrival, but it is possible to exclude potential users by erecting toll booths [1].

These characteristics fundamentally distinguish public goods from private goods and challenge the assumptions underlying the efficiency of competitive markets. In private good markets, individuals reveal their preferences through purchasing decisions, and market prices coordinate supply and demand efficiently. Public goods, however, lack both the revealing mechanism of prices and the excludability necessary to prevent free-riding, which ibs something we will delve later into this paper. This in turn creates the central economic problem: public goods tend to be undersupplied in market economies.

\vspace{1 cm}
\begin{table}[h!]
    \centering
    \caption{Exclusivity and Nonrivalry}
    \label{tab:public_private_goods_wrapped}
    % Define the total width of the table (e.g., \textwidth)
    \begin{tabularx}{\textwidth}{llXX}
        \toprule
        & & \multicolumn{2}{c}{\textbf{Exclusive}} \\
        \cmidrule(lr){3-4}
        & & \textbf{Yes} & \textbf{No} \\
        \midrule
        \multirow{2}{*}{\textbf{Rival}} & \textbf{Yes} & Hot dogs, automobiles, houses & Fishing grounds, public grazing land, clean air \\
        \cmidrule(lr){2-4}
        & \textbf{No} & Bridges, swimming pools, satellite television transmission (scrambled) & National defense, mosquito control, justice \\
        \bottomrule
    \end{tabularx}
\end{table}
\vspace{1cm}

Ths strategic nature of public goods provision means that individuals' decisions are interdepent. Unlike private goods where one's consumption is independent of others, in public goods settings,  each consumer derives a benefit from the
purchases of the other. This link between the consumers is absent with private
goods and introduces strategic interaction into the decision processes. With the strategic
interaction the consumers are involved in a game, so equilibrium is found using the
concept of a Nash equilibrium and this is where game theory comes into play.
\subsection{Why Public Goods Matter}

The characteristics of public goods create deviations from the conditions required for competitive markets to achieve Pareto efficiency. In private good markets, prices reveal preferences and coordinate supply and demand efficiently. But with public goods, the nonexcludability prevents the price mechanism from functioning properly, while nonrivalry means that the efficient allocation differs fundamentally from private goods.

This paper proceeds as follows: Section 2 establishes the mathematical framework for analyzing public goods provision. Section 3 demonstrates market failure through Nash equilibrium analysis. Section 4 examines the free rider problem in detail. Section 5 evaluates theoretical solutions including Lindahl pricing and Clarke-Groves mechanisms. Sections 6-7 analyze voting mechanisms and political economy models.

\section{Mathematical Foundations}

\subsection{General Equilibrium Model Setup}

Consider an economy with two individuals, A and B. This simplifies the mathematical buildup and makes it easier to approach the general solution.. Each individual has a utility function depending on consumption of a public good \(x\) and a private good: \(U_A(x, y_A)\) and \(U_B(x, y_B)\).

Labor is the only input in this economy. Person A has endowment \(l_A\) units of labor. These units of labor can be allocated to producing the public good \(l_A^x\) or the private good \(l_A^y\), subject to:
\begin{equation}
l_A^x + l_A^y = l_A
\end{equation}

Person B faces a similar constraint with endowment \(l_B\). The total labor endowment is \(l = l_A + l_B\).

\subsection{Production Functions}

Public good production depends on combined labor inputs from both individuals:
\begin{equation}
x = f(l_A^x + l_B^x)
\end{equation}

This equation captures the essential nature of \(x\) as a public good. Nonexclusivity is reflected by the fact that as A's labor input increases the amount of \(x\) that B consumes, A cannot prevent B from enjoying the fruits of A's labor, and vice versa. Nonrivalry is reflected by the fact that consumption of \(x\) by each person equals the total amount of \(x\) produced and A's consumption does not diminish what B can consume.

Private good production depends on separate labor inputs:
\begin{equation}
y_A = g(l_A^y), \quad y_B = g(l_B^y)
\end{equation}

\subsection{Efficiency Conditions: The Samuelson Rule}

To find the socially efficient outcome, we maximize one person's utility (say A's) for a given level of B's utility. The Lagrangian in this case will be:
\begin{equation}
\mathcal{L} = U_A(f(\cdot), g(l_A^y)) + \lambda[U_B(f(\cdot), g(l_B^y)) - \bar{U}]
\end{equation}

where \(\bar{U}\) is a constant representing B's utility level.

Taking first-order conditions with respect to \(l_A^y\) and \(l_B^y\) and solving yields:
\begin{equation}
\frac{\partial U_A/\partial x}{\partial U_A/\partial y_A} + \frac{\partial U_B/\partial x}{\partial U_B/\partial y_B} = \frac{g'}{f'}
\end{equation}

This is simply:
\begin{equation}
\text{MRS}_A + \text{MRS}_B = \text{RPT}
\label{eq:samuelson}
\end{equation}

This is the \textbf{Samuelson condition} by Paul Samuelson in 1954. It states that the sum of each individual's marginal rate of substitution (MRS) between the public and private good should equal the rate of change for product cost (the marginal cost(MC)).

The intuition behind it is as follows. For Public goods, the MRS in consumption must be the total amount of the private good that \textit{all} consumers would collectively be willing to give up to get one more unit of \(x\). This is because everyone obtains the benefits of extra \(x\) output. It contrasts sharply with private goods, where each consumer's MRS is equated individually to the price ratio.

\subsection{Vertical Summation of Demand}

The Samuelson condition states that demand curves for public goods must be summed \textit{vertically} rather than horizontally. The socially optimal price  for any quantity of public good equals to the sum of what each individual would pay . This simply means their combined marginal valuations

Thus if there are \(n\) consumers with individual inverse demand curves \(P_i(Q)\). The aggregate demand for a public good will  be:
\begin{equation}
P(Q) = \sum_{i=1}^{n} P_i(Q)
\end{equation}

This vertical summation reflects that all consumers simultaneously benefit from each unit of the public good.

\section{Market Failure Under Competition}

\subsection{Why Competitive Markets Fail}

Production of goods \(x\) and \(y\) in competitive markets will fail to achieve the efficiency condition in the samuelson Equation~\eqref{eq:samuelson}. When prices are perfectly competitive at \(p_x\) and \(p_y\). Each individual will equate his or her MRS to the price ratio \(p_x/p_y\). Profit maximization by producers leads to equality between the rate of product transformation and the price ratio:
\begin{equation}
\text{MRS}_A = \text{MRS}_B = \frac{p_x}{p_y} = \text{RPT}
\end{equation}

This violates the Samuelson condition. The price ratio \(p_x/p_y\) is too low and it provides insufficient incentive to produce \(x\). In the private markets, a consumer doesn't take into account how their spending on the public good benefits others. Thus, they devote too few resources to its production.

The allocation failure can be because of how private markets sum individual demands. The market demand curve reports marginal valuation for a \textit{single} consumer. For public goods, the value of producing one more unit is the \textit{sum} of each consumer's valuation as all benefit simultaneously.

\subsection{Inefficiency of Nash Equilibrium}

Generally, independent decision making fails to produce efficiency. This happens even with strategic agents who understand the game structure.
Consider the voluntary contribution mechanism where each agent \(i\) chooses contribution \(g_i\) to maximize:
\begin{equation}
\Pi_i(g_i, g_{-i}) = U_i\left(M_i - g_i, \sum_{j=1}^{n} g_j\right)
\end{equation}

where \(M_i\) is agent \(i\)'s income and \(g_{-i}\) represents all other agents' contributions.

A \textbf{Nash equilibrium} is a contribution profile \((g_1^*, \ldots, g_n^*)\) where:
\begin{equation}
\Pi_i(g_i^*, g_{-i}^*) \geq \Pi_i(g_i, g_{-i}^*) \quad \forall g_i \geq 0, \; \forall i
\end{equation}

For an interior solution where \(g_i^* > 0\), the first-order condition is:
\begin{equation}
-\frac{\partial U_i}{\partial x_i} + \frac{\partial U_i}{\partial G} = 0
\end{equation}

This implies that each agent equates their \textit{private} marginal benefit to the marginal cost i.e PMB = PMC. This means the agent ignores benefits that spill over to others and makes sure their benefits equal costs. With \(n\) agents, individual \(i\) only captures \(1/n\) of the total benefit when the good is enjoyed equally. This leads to systematic underprovision.

The public-good problem worsens as the number of agents increases. In the limit as \(n \to \infty\), any one person may provide essentially none of the public good, becoming a pure free rider.

\section{The Free Rider Problem}
The free rider problem is a market failure that occurs when individuals benefit from a public good or service without paying for it. It leads to under-provision because there is little incentive for people to contribute and thus it becomes a failure.
\subsection{The Roommates Dilemma: A Numerical Example}

To understand the free rider problem, let us consider two roommates A and B. Both have identical Cobb-Douglas preferences:
\begin{equation}
U_A(x, y_A) = x^{1/3} y_A^{2/3}
\label{eq:roommate_utility}
\end{equation}

where \(y_A\) is hours A spends reading (in this case will be the private good) and \(x = x_A + x_B\) is total hours spent cleaning the room (the public good). Each roommate has 10 hours per week available, so the budget constraint is:
\begin{equation}
x_i + y_i = 10
\end{equation}

\subsubsection{Nash Equilibrium Calculation}

Roommate A maximizes utility with repect to B's contribution \(x_B\). Then the Lagrangian is:
\begin{equation}
\mathcal{L} = (x_A + x_B)^{1/3} y_A^{2/3} + \lambda(10 - x_A - y_A)
\end{equation}

The first-order conditions:
\begin{align}
\frac{\partial \mathcal{L}}{\partial x_A} &= \frac{1}{3}(x_A + x_B)^{-2/3} y_A^{2/3} - \lambda = 0 \\
\frac{\partial \mathcal{L}}{\partial y_A} &= \frac{2}{3}(x_A + x_B)^{1/3} y_A^{-1/3} - \lambda = 0
\end{align}

Solving for y by equating (15) \& (16) using \(\lambda\) gives:
\begin{equation}
y_A = 2(x_A + x_B)
\end{equation}

By substituting (17) into the budget constraint would yield:
\begin{equation}
10 = x_A + y_A = x_A + 2(x_A + x_B) = 3x_A + 2x_B
\end{equation}

By symmetry, \(x_A = x_B = x^*\). This is in equilibrium. Therefore:
\begin{equation}
10 = 3x^* + 2x^* = 5x^* \implies x^* = 2
\end{equation}

Each roommate contributes 2 hours to cleaning. The total cleaning will be \(x = 4\) hours. The Equilibrium utilities for both roommates will be:
\begin{equation}
U_A = U_B = 4^{1/3} \cdot (20/3)^{2/3} \approx 6.3
\end{equation}

\subsubsection{Efficient Allocation}

For Cobb-Douglas preferences, the marginal rate of substitution will be:
\begin{equation}
\text{MRS}_A = \frac{\partial U_A/\partial x}{\partial U_A/\partial y_A} = \frac{y_A}{2x}
\end{equation}

From earlier we can also state that the Samuelson condition requires:
\begin{equation}
\text{MRS}_A + \text{MRS}_B = \frac{y_A}{2x} + \frac{y_B}{2x} = 1
\end{equation}

By symmetry, \(y_A = y_B = y^*\), so:
\begin{equation}
\frac{2y^*}{2x^*} = 1 \implies y^* = x^*
\end{equation}

Using the combined budget constraint \(20 = x^* + y_A + y_B = 3x^*\), we get:
\begin{equation}
x^* = \frac{20}{3} \approx 6.67, \quad x_A = x_B = \frac{10}{3} \approx 3.33
\end{equation}

Thus the efficient utilities for both roommates are are:
\begin{equation}
U_A = U_B = \left(\frac{20}{3}\right)^{1/3} \left(\frac{20}{3}\right)^{2/3} = \frac{20}{3} \approx 6.67
\end{equation}

\subsubsection{Comparison and the Free Rider Problem}

The Nash equilibrium suggests \textit{too little} cleaning. 2 hours each compared to 3.3 hours each in the efficient allocation. The gap comes because each roommate ignores the benefit the other roommate will get by them cleaning. Utility at Nash equilibrium (6.3) is lower than at the efficient allocation (6.67).
 The utility loss per roommate is:
\begin{equation}
\text{Utility Loss} = U^* - U^{NE} = 6.67 - 6.35 = 0.32 \text{ (approximately 4.8\% reduction)}
\end{equation}

This is the cost of free-riding: individually rational behavior produces collectively inferior outcomes.

This is comparable to the Prisoner's Dilemma. The Nash equilibrium (both contribute little) is Pareto dominated by another outcome (both contribute more).  yet neither has any incentive to deviate. 

\subsection{Example 2: Three People in a Dorm}

\subsubsection{3 player game}

Now consider a dormitory with three residents A, B, and C. All three with identical Cobb-Douglas preferences:
\begin{equation}
U_i(x, y_i) = x^{1/3} y_i^{2/3}, \quad i \in \{A, B, C\}
\end{equation}

Each has 10 hours available. The total cleaning hours will be denoted by: \(x = x_A + x_B + x_C\).

\subsubsection{Nash Equilibrium}

We can find th enash equilibrium by using symmetry. As each resident faces the same best response. Let us look at how resident A's will optimize their output:
\begin{center}
 Maximize \((x_A + x_B + x_C)^{1/3} y_A^{2/3}\) subject to \(x_A + y_A = 10\)   
\end{center}

The best response function will then be:
\begin{equation}
y_A = 2(x_A + x_B + x_C) \implies x_A = \frac{10 - 2(x_B + x_C)}{3}
\end{equation}

Again by symmetry in equilibrium \(x_A = x_B = x_C = x^*\), we will get:  
\begin{equation}
x^* = \frac{10 - 2(2x^*)}{3} = \frac{10 - 4x^*}{3} \implies 3x^* = 10 - 4x^* \implies 7x^* = 10 \implies x^* = \frac{10}{7} \approx 1.43
\end{equation}

Each person contributes approximately 1.43 hours, with total cleaning \(x = \frac{30}{7} \approx 4.29\) hours. Leisure is \(y_i = 10 - \frac{10}{7} = \frac{60}{7} \approx 8.57\) hours.

\textbf{Nash Equilibrium Utilities:}
\begin{equation}
U_i^{NE} = \left(\frac{30}{7}\right)^{1/3} \cdot \left(\frac{60}{7}\right)^{2/3}
\end{equation}

Computing numerically:
\begin{equation}
U_i^{NE} \approx 4.29^{1/3} \cdot 8.57^{2/3} \approx 1.626 \cdot 4.70 \approx 6.12
\end{equation}

\subsubsection{Efficient Allocation}

The Samuelson condition with three identical agents is:
\begin{equation}
\text{MRS}_A + \text{MRS}_B + \text{MRS}_C = \frac{y_A + y_B + y_C}{2x} = 1
\end{equation}

By symmetry, \(y_A = y_B = y_C = y^*\) and \(x_A = x_B = x_C\) at efficiency. The combined budget constraint is:
\begin{equation}
x + 3y = 30
\end{equation}

With \(y = x\) from the Samuelson condition:
\begin{equation}
x + 3x = 30 \implies x^* = \frac{30}{4} = 7.5
\end{equation}

Each person contributes \(x_i = 2.5\) hours, with leisure \(y^* = 7.5\) hours.

\textbf{Efficient Utilities:}
\begin{equation}
U_i^* = (7.5)^{1/3} \cdot (7.5)^{2/3} = 7.5 \approx 7.50
\end{equation}

\subsubsection{Comparison}

The inefficiency worsens with group size:

\begin{table}[h]
\centering
\caption{Nash vs. Efficient Provision: Two vs. Three Persons}
\label{tab:group_size}
\begin{tabular}{lcccc}
\toprule
& \textbf{Group Size} & \textbf{Nash } \(x_i^*\) & \textbf{Efficient } \(x_i^*\) & \textbf{Utility Gap} \\
\midrule
Two persons & 2 & 2.0 & 3.33 & 6.35 vs. 6.67 (-4.8\%) \\
Three persons & 1.43 & 2.5 & 6.12 vs. 7.50 (-18.4\%) \\
\bottomrule
\end{tabular}
\end{table}

When there are three people, each person's private benefit from their contribution is only \(1/3\) of the total social benefit. This weaker personal incentive leads to more severe underprovision (1.43 vs. 2.5 hours). The utility loss grows to 18.4\%.

\subsection{The Free Rider Problem with Many Agents}

As seen in the previous example, the problem intensifies as the number of agents grows. With \(n\) roommates, symmetric equilibrium requires:
\begin{equation}
10 = x_A + y_A = x_A + 2nx_A = (2n+1)x_A
\end{equation}

Thus:
\begin{equation}
x_A = \frac{10}{2n+1}
\end{equation}

As \(n \to \infty\), each roommate's contribution approaches zero. Yet the efficient contribution remains constant at \(x_A = 10/3\) per person, independent of \(n\).

This result demonstrates why large groups often fail to provide public goods voluntarily. A fraternity or large household lives in messy conditions not because individuals are lazy, but because they are rational. In the Nash equilibrium case, it is best for them to free-ride.

\section{Mechanisms for Solving the Public Goods Problem}

\subsection{Lindahl Pricing}

Erik Lindahl proposed an elegant conceptual solution in the 1920s[1]. Lindahl suggested individuals might voluntarily consent to be taxed for beneficial public goods if they knew others were also being taxed. Here tax shares reflect individual benefits.

\subsubsection{The Lindahl Mechanism}

Each individual \(i\) is quoted a personalized tax share \(\alpha_i\). It represents their proportion of the public good's cost. Individual A chooses the level of public good \(x\) to maximize:
\begin{equation}
U_A(x, g(l_A - \alpha_A f(x)))
\end{equation}

The first-order condition is:
\begin{equation}
\frac{\partial U_A/\partial x}{\partial U_A/\partial y_A} = \alpha_A \cdot \frac{g'}{f'} = \alpha_A \cdot \text{RPT}
\end{equation}

Similarly, individual B chooses \(x\) and satisfies the following:
\begin{equation}
\text{MRS}_B = \alpha_B \cdot \text{RPT}
\end{equation}

A \textbf{Lindahl equilibrium} occurs where \(\alpha_A + \alpha_B = 1\)—tax contributions precisely fund the public good. At this equilibrium:
\begin{equation}
\text{MRS}_A + \text{MRS}_B = (\alpha_A + \alpha_B) \cdot \text{RPT} = \text{RPT}
\end{equation}

This satisfies the Samuelson condition and thus achieves Pareto efficiency!

\subsubsection{Application to the Roommates}

If social convention suggests each roommate contributes half the cleaning effort (\(\alpha_A = \alpha_B = 1/2\)), each faces an effective price of 1/2 hour of personal effort per hour of total cleaning. With Cobb-Douglas preferences, each spends 1/3 of their 10-hour budget on cleaning at this price:
\begin{equation}
x_A = x_B = \frac{1}{3} \cdot 10 \cdot \frac{1}{1/2} = \frac{20}{3}
\end{equation}

Total cleaning is \(x = 20/3\), exactly the efficient level. This illustrates how appropriate personalized pricing achieves efficiency.

%\subsubsection{Practical Limitations}

\subsection{The Clarke-Groves (VCG) Mechanism}

Theodore Groves and Edward Clarke independently developed mechanisms that incentivize truthful preference revelation through clever tax schemes[1].

\subsubsection{Mechanism Design}

Suppose \(n\) individuals consider a public project. Each has private valuation \(v_i\). The government asks each to report valuation \(\hat{v}_i\), then:
\\
1. Implements the project if and only if \(\sum_i \hat{v}_i \geq 0\) \\
2. Charges each individual \(i\) a tax \(t_i\) if the project is implemented, it can be represented as:
\begin{equation}
t_i = \sum_{j \neq i} \hat{v}_j
\end{equation}

Otherwise, there will be a zero tax.

\subsubsection{Incentive Compatibility}

Individual \(i\)'s utility from reporting \(\hat{v}_i\) when true valuation is \(v_i\) is:
\begin{equation}
U_i(\hat{v}_i) = 
\begin{cases}
v_i - t_i = v_i - \sum_{j \neq i} \hat{v}_j & \text{if project implemented} \\
0 & \text{otherwise}
\end{cases}
\end{equation}

The project is implemented when \(\hat{v}_i + \sum_{j \neq i} \hat{v}_j \geq 0\). Note that \(\sum_{j \neq i} \hat{v}_j\) is independent of \(i\)'s report.\\

\textbf{Therefore:} \\
- When it is beneficial to \(i\), \(v_i > 0\). Thus, reporting truthfully will ensure the project proceeds   \\
- When it is harmful to \(i\), \(v_i < 0\). Reporting prevents implementation\\

Truthful revelation \(\hat{v}_i = v_i\) is a \textbf{dominant strategy}. It is optimal regardless of what other agents report.

\subsubsection{The Budget Deficit Problem}

Clarke-Groves mechanisms face a critical limitation. It typically runs a budget deficits even after achieving truthful revelation and efficiency.

Total revenue collected is:
\begin{equation}
R = \sum_{i=1}^{n} t_i = \sum_{i=1}^{n} \sum_{j \neq i} \hat{v}_j = (n-1) \sum_{i=1}^{n} v_i
\end{equation}

The project costs \(C\). It's implemented when \(\sum_i v_i \geq C\). The revenue is \((n-1) \sum_i v_i\). Unless \(n = 2\), revenue doesn't cover costs exactly and simply the mechanism runs a surplus or deficit.

This budget imbalance prevents full implementation of Clarke-Groves in practice. Redistributing surplus or financing deficits typically requires distortionary taxation. Which undermines efficiency gains.

\section{Voting and Resource Allocation}

Many public goods decisions are made through voting rather than markets. Understanding how voting mechanisms perform is crucial for analyzing real-world public provision.

\subsection{Majority Rule}

While commonly used, majority rule is not obviously optimal. Different institutions require different decisive proportions:\\
- Simple majority (50\%+1) for most legislation\\
- Two-thirds (67\%) for U.S. Constitutional amendments  \\
- Unanimous wins for some organizations (Quaker meetings)

Our analysis focuses on simple majority rule. The choice of threshold significantly affects outcomes.

\subsection{The Paradox of Voting}

In the 1780s, Condorcet discovered that majority rule may cycle endlessly without reaching equilibrium.

Consider three voters: Smith, Jones, and Fudd. Each choosing among three public good spending levels (A=low, B=medium, C=high) with preferences:

\begin{table}[h]
\centering
\begin{tabular}{lccc}
\toprule
& \textbf{Choice 1} & \textbf{Choice 2} & \textbf{Choice 3} \\
\midrule
Smith & A & B & C \\
Jones & B & C & A \\
Fudd & C & A & B \\
\bottomrule
\end{tabular}
\end{table}

Comparing options pairwise:\\
- A vs. B: A wins (Smith and Fudd favor A)\\
- A vs. C: C wins (Jones and Fudd favor C)\\
- C vs. B: B wins (Smith and Jones favor B)

We cycle: \(A \succ B \succ C \succ A\). No stable outcome exists. The final choice depends on agenda setting or when voting stops. These factors are unrelated to actual preferences.

\subsection{Single-Peaked Preferences and the Median Voter Theorem}

Duncan Black showed in 1948 that cycles can be avoided with restrictions on preferences. If preferences are \textbf{single-peaked} (having only one local maximum) then majority rule selects the median voter's preferred outcome.

Single-peaked means that as the policy variable (e.g., spending level) increases, utility first rises to a peak then declines This happens without multiple local maxima. In Condorcet's example, Fudd's preferences causes the cycle and has two peaks (A and C).

Now, instead Fudd had preferences peaked only at C, then option B would defeat both A and C by 2-to-1 votes. Option B is Jones's preferred choice, Jones is the median voter whose preferences lie between Smith (preferring A) and Fudd (preferring C).
\begin{figure}[htbp]
    \centering
    \includegraphics[width=0.8\textwidth]{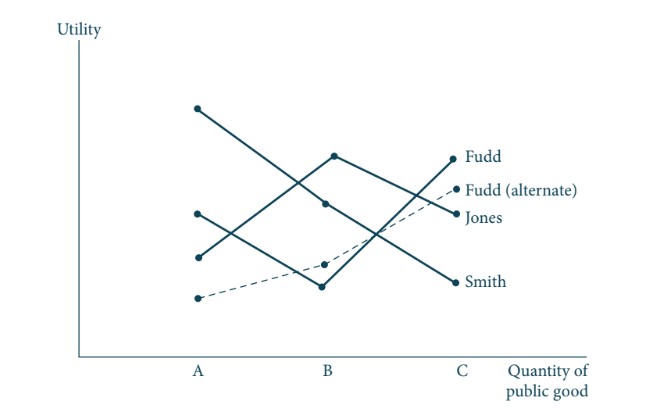}
    \caption{This figure illustrates the preferences in the Table above. Smith’s and Jones’s preferences are single peaked, but
Fudd’s have two local peaks and these yield the voting paradox. If Fudd’s preferences had instead been
single peaked (the dashed line), then option B would have been chosen as the preferred choice of the
median voter (Jones).}
    \label{Single-Peaked
Preferences and the
Median Voter Theorem}
\end{figure}

\textbf{Median Voter Theorem}: With single-peaked preferences over a one-dimensional policy space, the majority rule selects the preferred policy of the median voter. This theorem is foundational for political economy models. It predicts that political competition drives policy toward the median voter's preferences.

\section{A Political Economy Model}

\subsection{Setup}

Consider a community with \(n\) voters, each with income \(y_i\). Utility depends on private consumption \(c_i\) and a public good \(g\):
\begin{equation}
U_i(c_i, g) = c_i + f(g)
\end{equation}

where \(f' > 0\) and \(f'' < 0\) (concave benefits from public good).

The public good is financed by proportional income taxation at rate \(t\), so:
\begin{equation}
c_i = (1-t) y_i
\end{equation}

The government budget constraint is:
\begin{equation}
g = t \sum_{j=1}^{n} y_j = t n \bar{y}
\end{equation}

where \(\bar{y}\) is average income.

\subsection{Individual Preferences}

Each individual chooses preferred \(g\) to maximize:
\begin{equation}
U_i(g) = \left(1 - \frac{g}{n\bar{y}}\right) y_i + f(g) = y_i - \frac{y_i}{n\bar{y}} g + f(g)
\end{equation}

First-order condition:
\begin{equation}
f'(g_i^*) = \frac{y_i}{n\bar{y}}
\end{equation}

This shows that desired spending on \(g\) is \textit{inversely} related to income. High-income voters pay more in taxes (proportional to income) but receive the same benefits as low-income voters. Thus, high-income voters prefer less public spending.

\subsection{Median Voter Equilibrium}

Under majority rule, the level of \(g\) chosen is that preferred by the voter with median income \(y_m\):
\begin{equation}
f'(g^*) = \frac{y_m}{n\bar{y}}
\end{equation}

Since income distributions are typically right-skewed implying  \(y_m < \bar{y}\). This implies:
\begin{equation}
f'(g^*) = \frac{y_m}{n\bar{y}} < \frac{\bar{y}}{n\bar{y}} = \frac{1}{n}
\end{equation}

The more unequal the income distribution (larger gap between \(y_m\) and \(\bar{y}\)), the higher the equilibrium tax rate and public spending.
\section{Conclusion}

Public goods present fundamental challenges to market-based resource allocation. The combination of nonexclusivity and nonrivalry creates incentives for free-riding. This leads competitive markets and strategic voluntary contribution games to undersupply public goods relative to Pareto-efficient levels.


\begin{thebibliography}{1}
\bibitem{Nicholson} Nicholson, W., \& Snyder, C. (2017). \textit{Microeconomic theory: Basic principles and extensions}. Cengage Learning.

\bibitem{Gruber} Gruber, J. (2016). \textit{Public finance and public policy}. Worth Publishers.

\bibitem{Taehyon Choi} Choi, T., \& Robertson, P. (2019) 
\textit{ Contributors and Free-Riders in Collaborative Governance: A Computational Exploration of Social Motivation and Its Effects.} Journal of Public Administration Research and Theory. https://doi.org/10.1093/jopart/muy068

\end{thebibliography}
\end{document}